\pdfoutput=1
\documentclass[pra,twocolumn,superscriptaddress,longbibliography]{revtex4-2}
\usepackage{graphicx,empheq,amssymb,scalerel,tikz}
\usepackage[colorlinks=true, citecolor=blue, urlcolor=blue, linkcolor=blue ]{hyperref}
\usetikzlibrary{svg.path}
\definecolor{orcidlogocol}{HTML}{A6CE39}
\tikzset{
	orcidlogo/.pic={
		\fill[orcidlogocol] svg{M256,128c0,70.7-57.3,128-128,128C57.3,256,0,198.7,0,128C0,57.3,57.3,0,128,0C198.7,0,256,57.3,256,128z};
		\fill[white] svg{M86.3,186.2H70.9V79.1h15.4v48.4V186.2z}
		svg{M108.9,79.1h41.6c39.6,0,57,28.3,57,53.6c0,27.5-21.5,53.6-56.8,53.6h-41.8V79.1z M124.3,172.4h24.5c34.9,0,42.9-26.5,42.9-39.7c0-21.5-13.7-39.7-43.7-39.7h-23.7V172.4z}
		 svg{M88.7,56.8c0,5.5-4.5,10.1-10.1,10.1c-5.6,0-10.1-4.6-10.1-10.1c0-5.6,4.5-10.1,10.1-10.1C84.2,46.7,88.7,51.3,88.7,56.8z};}}
\newcommand\orcid[1]{\href{https://orcid.org/#1}{\mbox{\scalerel*{\begin{tikzpicture}[yscale=-1,transform shape]\pic{orcidlogo};\end{tikzpicture}}{|}}}}

\begin{document}
\title{Non-Markovian quantum interconnect formed by a surface plasmon polariton waveguide}
\author{Chun-Jie Yang\orcid{0000-0003-2137-6958}}
\affiliation{School of Physics, Henan Normal University, Xinxiang 453007, China}
\author{Xin-Yue Liu}
\affiliation{Key Laboratory of Quantum Theory and Applications of MoE, Lanzhou University, Lanzhou 730000, China}
\affiliation{Lanzhou Center for Theoretical Physics, Key Laboratory of Theoretical Physics of Gansu Province, Lanzhou University, Lanzhou 730000, China}
\author{Shi-Qiang Xia}\email{xiashiqiang@htu.edu.cn}
\affiliation{School of Physics, Henan Normal University, Xinxiang 453007, China}
\author{Si-Yuan Bai}
\affiliation{Key Laboratory of Quantum Theory and Applications of MoE, Lanzhou University, Lanzhou 730000, China}
\affiliation{Lanzhou Center for Theoretical Physics, Key Laboratory of Theoretical Physics of Gansu Province, Lanzhou University, Lanzhou 730000, China}
\author{Jun-Hong An\orcid{0000-0002-3475-0729}}\email{anjhong@lzu.edu.cn}
\affiliation{Key Laboratory of Quantum Theory and Applications of MoE, Lanzhou University, Lanzhou 730000, China}
\affiliation{Lanzhou Center for Theoretical Physics, Key Laboratory of Theoretical Physics of Gansu Province, Lanzhou University, Lanzhou 730000, China}

\begin{abstract}
Allowing the generation of effective interactions between distant quantum emitters (QEs) via flying photons, quantum interconnect (QI) is essentially a light-matter interface and acts as a building block in quantum technologies. A surface plasmon polariton (SPP) supported by a metallic waveguide provides an ideal interface to explore strong light-matter couplings and to realize QI. However, the loss of SPP in metal makes the mediated entanglement of the QEs damp with the increase of the distance and time, which hinders its applications. We propose a scheme of non-Markovian QI formed by the SPP of a metallic nanowire. A mechanism to make the generated entanglement of the QEs persistent is discovered. We find that, as long as bound states are formed in the energy spectrum of total QE-SPP system, the damping of the SPP-mediated entanglement is overcome even in the presence of the metal absorption to the SPP. Our finding enriches our understanding of light-matter couplings in absorptive medium and paves the way for using the SPP in designing QI.
\end{abstract}
\maketitle

\section{Introduction}
Pursuing the utilization of quantum resources to make scientific and technological innovations, a quantum revolution is in the making \cite{doi:10.1098/rsta.2003.1227,PRXQuantum.1.020101}. In the past decades, dramatic progress has been made in individual quantum systems, including trapped ions, neutral atoms, superconducting qubits, and photonics \cite{Graham2022,Pino2021,Mirhosseini2020,doi:10.1126/science.abe8770}. An urgent requirement is to integrate them to form advanced technologies that bring essential benefits \cite{doi:10.1126/science.abk2617,doi:10.1126/sciadv.aba4508,Kimble2008}. One promising solution is quantum interconnect (QI), which is a functional device to coherently link and distribute entanglement among systems across different length scales \cite{PRXQuantum.2.017002}. The core of a QI is a finely controlled light-matter interface to correlate distant matters via flying photons \cite{Kurpiers2018,doi:10.1126/science.1103346,RevModPhys.87.1379,Axline2018,PhysRevLett.125.260502,PhysRevLett.120.200501,PhysRevLett.124.110501}. Efficient QI among quantum emitters (QEs) has been explored in different kinds of waveguide quantum electrodynamics (QED) systems \cite{PhysRevLett.118.133601,PhysRevLett.124.063602,doi:10.1126/sciadv.abb8780,Kannan2023,Kannan2020,Corzo2019,PhysRevLett.127.173601,PhysRevLett.127.273602,RevModPhys.95.015002,doi:10.1126/sciadv.aaw0297,Zanner2022,Niu2023}.

\begin{figure}[t]
\includegraphics[width=0.9\columnwidth]{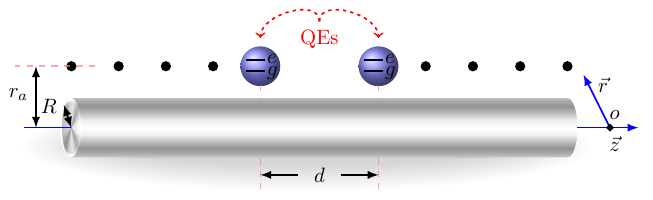}
\caption{Scheme of a QI formed by a SPP waveguide to entangle an array of $N$ QEs positioned at a distance $r_a$ from the axis of a metallic nanowire with radius $R$. The QEs are modeled as two-level systems and separated by a distance $d$. }\label{Fig1}
\end{figure}

A surface plasmon polariton (SPP) supported by a metallic waveguide supplies an ideal platform to realize QI. SPP is a hybrid excitation mode of electromagnetic field (EMF) and electron density waves on the metal-dielectric interface. Confining EMF in a spatial scale far below the diffraction limit, SPP has attracted an extensive interest to explore strong quantized light-matter couplings \cite{Santhosh2016,PhysRevLett.124.063902,PhysRevLett.114.036802,Cunningham2019,PhysRevLett.126.223603,Goncalves2020,PhysRevLett.117.107401,PhysRevLett.110.126801} and led to fascinating applications in quantum technologies \cite{Barnes2003,Tame2013,doi:10.1021/acs.chemrev.0c01028,doi:10.1126/sciadv.abn2026,Chang2007,doi:10.1021/acs.nanolett.9b01137,doi:10.1021/acs.nanolett.6b00706,You2021,doi:10.1126/sciadv.aau8763,doi:10.1021/acs.nanolett.1c02495,Dhama2022}. It has been found that SPP can act as a quantum bus to mediate coherent interactions and generate entanglement among separated QEs \cite{PhysRevB.84.045310,PhysRevB.84.235306,PhysRevLett.106.020501,PhysRevB.106.155409,Ryom2022,PhysRevA.106.012402,PhysRevB.84.235306,PhysRevLett.106.020501,PhysRevB.106.155409,acsphotonics,doi:10.1021/acsphotonics.7b00717,C9NR05083C,PhysRevB.96.075405}, which exhibits an attractive prospect of realizing QI by SPP. However, the metal absorption to the EMF makes the SPP short-lived, which severely restricts its practical applications \cite{doi:10.1126/science.aax3766}. We really see that the SPP-induced entanglement between QEs is dynamically transient and tends to vanish in the long-time limit and with increasing the QE distance in the absence of coherent driving \cite{PhysRevB.84.235306,PhysRevLett.106.020501,PhysRevB.106.155409,acsphotonics,doi:10.1021/acsphotonics.7b00717,C9NR05083C,PhysRevB.96.075405}. In practice, one generally desires that a persistent entanglement of the distant QEs could be mediated by QIs. Thus, how to suppress the destructive influence of the lossy SPPs on the QEs is a key problem in applying SPPs in QI.

Here, we propose a scheme of non-Markovian QI to distribute persistent entanglement among an array of distant QEs by the SPP propagating along the surface of a metallic nanowire. We find that the subdiffraction feature of the SPP endows our system with the strong light-matter coupling nature. It causes that the $N$ QEs are so hybridized with the SPP that at most $N$ QE-SPP bound states are formed. This mechanism makes the entanglement among the QEs mediated by the SPP persistently preserved in the steady state, which differs from the asymptotic vanishing in the Born-Markovian approximation \cite{PhysRevB.84.045310,PhysRevB.84.235306,PhysRevLett.106.020501,Ryom2022,PhysRevA.106.012402,PhysRevB.106.155409,PhysRevLett.106.020501,PhysRevB.106.155409,C9NR05083C,PhysRevB.96.075405,PhysRevB.82.075427}. Solving the decoherence problem of the QEs caused by the lossy SPP, the result enriches our understanding on the light-matter interactions in absorptive media and paves the way for applying SPPs in realizing QI to scalable quantum devices.

\section{Model and Hamiltonian}\label{part1}
Nanowires supporting propagating SPPs with higher field confinement can be used as one-dimensional nanowaveguide to connect QEs in separated positions, providing fundamental building blocks for nanophotonic integrated circuits \cite{doi:10.1021/acs.nanolett.7b05448,doi:10.1021/acs.chemrev.7b00441}. However, the loss of SPPs in metal hinders its applications. Previous works have shown that, going beyond the weak-coupling regime, once the QE is placed very close to the surface of plasmonic nanostructures, the SPP-induced dissipation can be suppressed in the strong coupling regime \cite{PhysRevB.95.075412,PhysRevB.95.161408}. Inspired by this research, we are here interested in the realization of QI by coupling QEs to a plasmonic waveguide going beyond the Born-Makrovian approximation.

We consider a system consisting of an array of $N$ QEs coupled to a cylindrical metal nanowire (see Fig. \ref{Fig1}). Being positioned on a line at a distance $r_a$ from the nanowire axis, the QEs are equally separated in a distance $d$ and modeled by two-level systems with frequency $\omega_0$. The electric permittivity of the nanowire is described by the Drude model as $\varepsilon _\text{m}(\omega )=\varepsilon _{\infty }-\omega _{p}^{2}/[\omega(\omega +i\gamma _{p})]$, where $\varepsilon _{\infty }$ is the high-frequency limit of $\varepsilon _\text{m}(\omega )$, $\omega _{p}$ is the bulk plasma frequency, and $\gamma _{p}$ is a damping factor of the EMF in the metal. Rather than coupling with the localized SPPs around a metal nanoparticle \cite{PhysRevResearch.1.023027}, the transmission characteristic of SPPs along the nanowire endows the system with superiorities in connecting QEs over large distance, which is crucial in designing QI.

Three modes are triggered by the radiation field of the QEs. The first one is the radiation mode to the free space. The second one is the non-radiative mode absorbed by the metal. The last one is the SPP, i.e., a hybrid mode of the EMF and the electron density waves, propagating along the nanowire. The SPP enables a confinement of the EMF beyond the diffraction limit on the metal-space interface, which makes the system an ideal platform to realize the strong light-matter interactions. To describe the quantum features of such couplings, a quantization of the EMF in the absorptive medium is needed. A scheme of macroscopic quantum electrodynamics was developed using the Green's tensor \cite{PhysRevA.53.1818,PhysRevA.57.3931}, where the canonical commutation relations of the EMF are guaranteed by introducing a Langevin noise to describe the medium absorption. Then the quantized electric field is
\begin{equation}
\mathbf{\hat{E}}(\mathbf{r},\omega )=\frac{ic^{-2}\omega ^{2}}{\sqrt{\pi \varepsilon _{0}/\hbar }}\int d^{3}\mathbf{r}^{\prime }\sqrt{\text{Im}[\varepsilon _{\text{m}}(\omega )]}\mathbf{G(r},\mathbf{r}^{\prime},\omega )\cdot\mathbf{\hat{f}}(\mathbf{r}^{\prime },\omega ),\label{drm}
\end{equation}
where $c$ is the speed of light, $\varepsilon _{0}$ is the vacuum permittivity, and $\mathbf{\hat{f}}(\mathbf{r},\omega )$ obeying $[\mathbf{\hat{f}(r},\omega),\mathbf{\hat{f}}^{\dag }(\mathbf{r}^{\prime},\omega ^{\prime })]=\delta (\mathbf{r}-\mathbf{r}^{\prime })\delta (\omega-\omega ^{\prime })$ is the annihilation operator of EMF on the metal-space interface. The Green's tensor $\mathbf{G(r},\mathbf{r}^{\prime },\omega )$ denotes the field in frequency $\omega $ at $\mathbf{r}$ triggered by a point source at $\mathbf{r}^{\prime }$ and is determined by $[{\pmb\nabla }\times {\pmb\nabla} \times -\omega ^{2}c^{-2}\varepsilon _\text{m}\left(\omega \right) ]\mathbf{G(r},\mathbf{r}^{\prime },\omega )=\mathbf{I}\delta (\mathbf{r}-\mathbf{r}^{\prime })$, where $\mathbf{I}$ is an identity matrix. The spatial distribution of the three modes has been incorporated in $\mathbf{G(r},\mathbf{r}^{\prime },\omega )$. Thus, via $\mathbf{G(r},\mathbf{r}^{\prime },\omega )$, we obtain a quantization of the three modes triggered by the EMF in the absorptive medium subject to the system geometry \cite{PhysRevB.82.075427}. Note that, widely used to study the light-matter interactions in absorptive media \cite{PhysRevLett.127.013602,PhysRevLett.127.250402,PhysRevLett.122.213901,PhysRevLett.128.167403,Lozano2023,doi:10.1073/pnas.1611924114,doi:10.1515/nanoph-2020-0451,Rivera2020}, the quantization method also applies to cases when the losses in the medium may be disregarded. A careful inspection indicates that, rather than vanishing, the integral makes the quantized electric-field operator reduce to the standard form in free space when $\epsilon_{\text{I}}(\omega)\rightarrow0$ (see Appendix \ref{Appen-quan}). $\mathbf{G(r},\mathbf{r}^{\prime },\omega )$ is analytically solvable for a cylindrical nanowire (see Appendix \ref{Appen-Green}). EMF also induces localized surface plasmons on a nanosphere, but they are not suitable for guiding wave in a long distance \cite{PhysRevResearch.1.023027}.

Under the dipole and rotating-wave approximations, the Hamiltonian describing the interactions between the QEs and the three modes reads
\begin{eqnarray}\label{Hamiltonian}
\hat{H} &=&\sum_{l=1}^{N}\hbar \omega _{0}\hat{\sigma}^{\dag}_{l}\hat{\sigma}_l+\int d^{3}\mathbf{r}\int_{0}^{\infty} d\omega \hbar \omega \mathbf{\hat{f}}^{\dag }(\mathbf{r},\omega )\cdot\mathbf{\hat{f}(r},\omega )  \nonumber \\
&&-\sum_{l=1}^{N}\int_{0}^{\infty} d\omega \lbrack \pmb{\mu} _{l}\cdot \mathbf{\hat{E}}(\mathbf{r}_{l},\omega )\hat{\sigma}_{l}^{\dag}+\textrm{H.c.}],
\end{eqnarray}
where $\hat{\sigma}_{l}=|g_{l}\rangle \langle e_{l}|$ is the transition operator from the excited state $|e_{l}\rangle $ to the ground state $| g_{l}\rangle$, $\pmb{\mu}_{l}$ is the dipole moment, and $\mathbf{r}_{l}$ is the coordinate of the $l$th QE. Equation \eqref{Hamiltonian} has been widely used to describe the interaction between QEs and the EMF in absorptive medium \cite{PhysRevA.62.053804,RevModPhys.95.015002,PhysRevLett.122.213901,PhysRevLett.127.013602}.

\section{Non-Markovian QI}\label{part2}
It is easy to verify that the total excitation number $\hat{\mathcal{N}}=\sum_{l}\hat{\sigma}_{l}^{\dag}\hat{\sigma}_{l}+\int d^3{\bf r}\int d\omega \mathbf{\hat{f}}^{\dag }(\mathbf{r},\omega )\cdot \mathbf{\hat{f}}(\mathbf{r},\omega )$ is conserved due to $[\hat{H},\hat{\mathcal{N}}]=0$. Thus, once being prepared in a single-excitation state initially, the state of the total system evolves to $|\Psi (t)\rangle =[\sum_{l}c_{l}(t)\hat{\sigma}_{l}^{\dag }+\int d^{3}\mathbf{r}\int d\omega c_{\mathbf{r},\omega }(t)\mathbf{\hat{f}}^{\dag }(\mathbf{r},\omega )]|G;\{0_{\mathbf{r},\omega }\}\rangle$, where $|G\rangle $ denotes that all the QEs are in their ground state and $|\{0_{\mathbf{r},\omega }\}\rangle $ is the vacuum state of the EMF. We can derive that the vector $\mathbf{c}(t)=\left( \begin{array}{ccc}  c_1(t) & \cdots & c_N(t) \\  \end{array}\right)^\text{T}$ formed by the excited-state probability amplitudes of the QEs obeys \cite{PhysRevResearch.1.023027}
\begin{equation}
\mathbf{\dot{c}}(t)+i\omega _{0}\mathbf{c}(t)+\int_{0}^{t}d\tau\int_{0}^{\infty }d\omega e^{-i\omega (t-\tau )}\mathbf{J}(\omega )\mathbf{c}(\tau )=0,  \label{evo-1}
\end{equation}
where $\mathbf{J}(\omega)$ is a $N$-by-$N$ matrix with element $J_{lj}(\omega) =\omega ^{2}\pmb{\mu} _{l}\cdot \textrm{Im}[\mathbf{G}(\mathbf{r}_{l},\mathbf{r}_{j},\omega )]\cdot \pmb{\mu} _{j}^{\ast }/(\pi\hbar\varepsilon_{0}c^2)$ characterizing the correlated spectral density between the $l$th and $j$th QEs. Such an correlation indicates that, although direct couplings between the QEs are absent in Eq. \eqref{Hamiltonian}, their indirect couplings are triggered via exchanging the SPPs. We consider that the QEs have identical dipole moments such that $J_{lj}(\omega)=J_{mn}(\omega)\equiv J_{|l-j|}(\omega)$ for $|l-j|=|m-n|$. Reflecting the memory effects, the convolution in Eq. \eqref{evo-1} renders the dynamics non-Markovian. It is \ prominent in quantum plasmonics because the light-matter interactions are significantly enhanced by the sub-diffraction confinement of the SPPs to EMF \cite{Thanopulos:19,PhysRevB.105.245411}. Under the Born-Markovian approximation, we obtain the solution of Eq. \eqref{evo-1} as $\mathbf{c}(t)=\exp [-(\bar{\pmb{\gamma}}/2+i\bar{\pmb{\omega}})t]\mathbf{c}(0)$ with $\bar{\pmb{\gamma}}=2\pi \mathbf{J}(\omega _{0})$ and $\bar{\pmb{\omega}}=\omega _{0}+\mathcal{P}\int_{0}^{\infty }d\omega \frac{\mathbf{J}(\omega )}{\omega -\omega _{0}}$, where $\mathcal{P}$ represents the Cauchy principle value. The positivity of the self-correlated spectral density $J_{0}(\omega_{0})$ ensures the existence of positive values in elements of $\bar{\pmb{\gamma}}$, which leads to a completely decay of the quantum coherence in long-time limit \cite{PhysRevB.82.075427}. As shown in the following, a distinct behavior appears beyond the Born-Markovian approximation, which needs an exact solution of the dynamics.  

\begin{figure*}[tbp]
\includegraphics[width=0.95\textwidth]{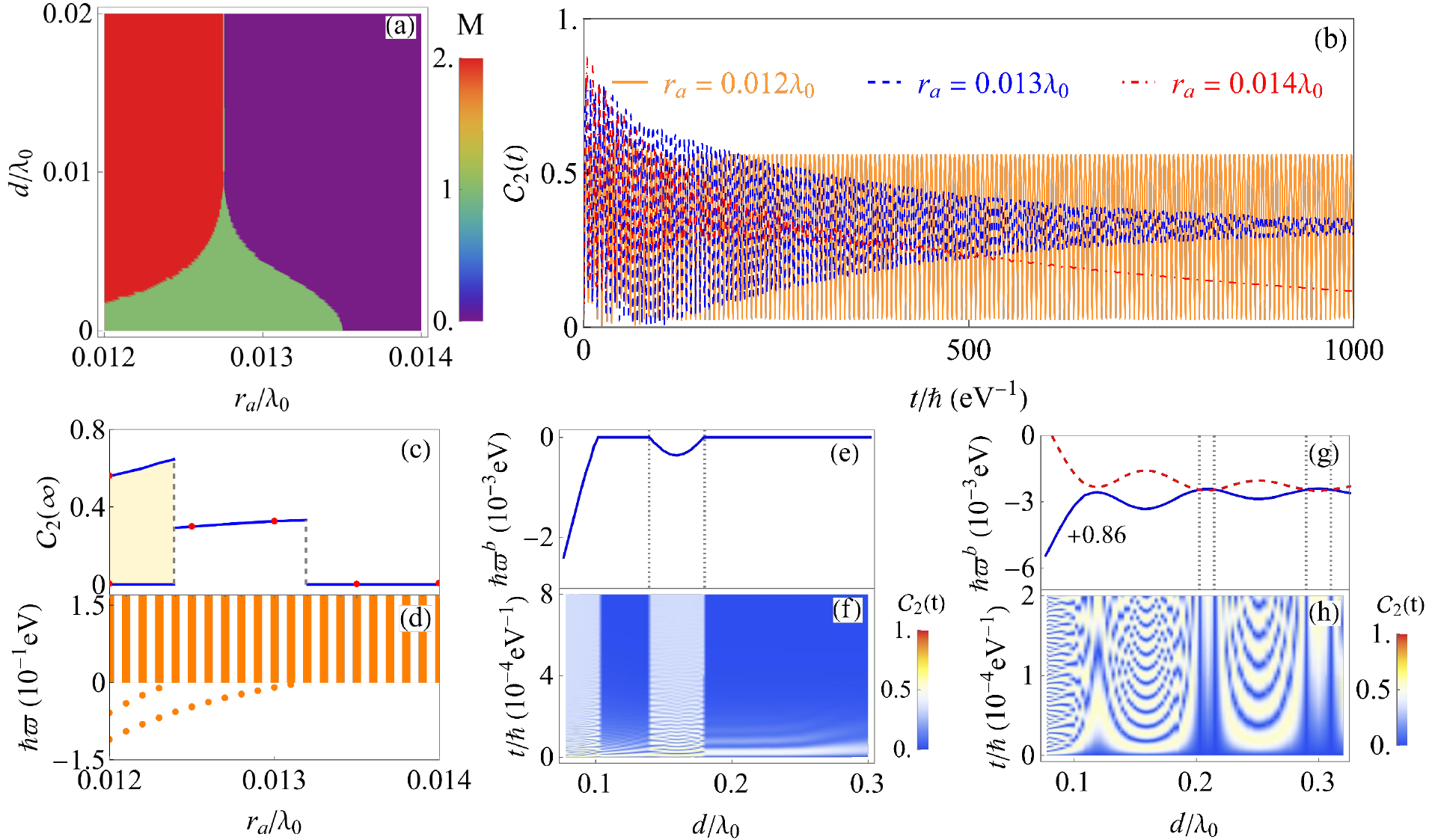}
\caption{(a) Numbers of the bound states. (b) Evolution and (c) long-time values of $\mathcal{C}_2(t)$ in different $r_a$ when $d=5$ nm by solving Eq. \eqref{evo-1}. In (c), the blue lines and red points are from Eq. \eqref{double-c} and Eq. \eqref{evo-1}, respectively, and the orange region covers the values of $\mathcal{C}_2(\infty )$ during its persistent oscillation. (d) Energy spectrum of the whole system, which consists of a continuous band and two branches of bound states in the band gap. Eigenenergies of the bound states and evolution of $\mathcal{C}_2(t)$ in different $d$ when $r_a=0.0126\lambda_0$ in (e) and (f) and $0.012\lambda_0$ in (g) and (h). The dotted lines highlight the critical points in which the entanglement is lost in the long-time limit. Other parameters are $\hbar \gamma_0=10^{-4}$ eV, $N=2$, $R=0.01\lambda_0$, $\omega_0=0.1\omega_p$, and $\hbar\omega_p=9$ eV.}
\label{Fig2}
\end{figure*}
Equation \eqref{evo-1} is formally solvable by the Laplace transform $\mathbf{\bar{c}}(s)=\int_{0}^{\infty }e^{-st}\mathbf{c}(t)dt$, which results in $\mathbf{\bar{c}}(s)=\big[ s+i\omega _{0}+\int_{0}^{\infty} d\omega \frac{\mathbf{J}(\omega )}{s+i\omega }\big] ^{-1}\mathbf{c}(0)$. $\mathbf{c}(t)$ is obtained via the inverse Laplace transform to $\mathbf{\bar{c}}(s)$, which can be done by finding the poles from ($\varpi =is$)
\begin{equation}\label{eigen-solution}
y_{j}(\varpi)\equiv\omega _{0}-\int_{0}^{\infty }d\omega \frac{D_{j}(\omega )}{\omega -\varpi }=\varpi,~(j=1,\cdots,N)
\end{equation}
with $D_{j}(\omega)$ being the $j$th eigenvalues of $\mathbf{J}(\omega)$. It is interesting to find that the root $\varpi$ multiplied by $\hbar$ is just the eigenenergy of Eq. \eqref{Hamiltonian} in the single-excitation subspace (see Appendix \ref{Appen-eigenenergy}). It indicates that the dynamics of the system characterized by ${\bf c}(t)$ is intrinsically determined by the features of the energy spectrum formed by $\hbar\varpi$. Because $y_{j}(\varpi)$ is a decreasing function in the region of $\varpi<0$, Eq. \eqref{eigen-solution} has a discrete root $\varpi^b_j$ provided $y_{j}(0)<0$. The QE-SPP eigenstate for the eigenenergy $\hbar\varpi^b_j$ falling in the bandgap regime of the SPPs is called a bound state, whose formation has profound consequences on the dynamics of the QEs \cite{PhysRevResearch.1.023027,PhysRevB.97.115402}. It is expected that, depending on the system parameters, $N$ bound states could be formed at most for our $N$-QE configuration. In the region of $\varpi>0$, $y(\varpi)$ is ill-defined and thus Eq. \eqref{eigen-solution} has an infinite number of roots, which form a continuous energy band. According to the Cauchy's residue theorem, we have
\begin{equation}
\mathbf{c}(t)=\mathbf{Z}(t)+\int_{0}^{\infty}\frac{d\varpi }{2\pi }[\mathbf{\bar{c}}( 0^{+}-i\varpi)-\mathbf{\bar{c}}(0^{-}-i\varpi)]e^{-i\varpi t},\label{sol}
\end{equation}
where $\mathbf{Z}(t)=-i\sum_{j=1}^M$Res$[\mathbf{\bar{c}}(-i\varpi _{j}^{b})]e^{-i\varpi _{j}^{b}t}$, with $\text{Res}[\cdot]$ being the residue contributed by the $j$th bound state $\varpi_{j}^{b}$ and $M$ being the number of formed bound states. The second term comes from the eigenstates of the energy band. Oscillating with time in continuously changing frequencies, the second term tends to zero in the long-time limit due to the out-of-phase interference. Thus, if the bound state is absent, then $\lim_{t\rightarrow\infty}\mathbf{c}(t)={\bf 0}$ characterizes a complete dissipation, while if the bound states are formed, then $\lim_{t\rightarrow\infty}\mathbf{c}(t)=\mathbf{Z}(t)$ implies a dissipation suppression. Such behaviors are absent in the Born-Markovian approximation and unique for the non-Markovian dynamics. This result inspires us to propose a scheme to realize a QI by the metal nanowire. Such a QI acting as a quantum bus permits us to distribute a stable entanglement among the QEs acting as nodes of a quantum network in the non-Markovian dynamics due to the formation of the bound states.

First, we consider $N=2$ under the initial condition $|\Psi(0)\rangle=|e_{1}g_{2}\rangle $. We calculate (see Appendix \ref{Appen-solution})
\begin{equation}
\mathbf{Z}(t)=\frac{1}{2}\sum_{j=1}^{M}K^{(j)}_{01}e^{-i\varpi _{j}^{b}t}
\left(
\begin{array}{c}
1 \\
(-1)^{j+1}%
\end{array}%
\right),
\end{equation}
with $K_{mn}^{(j)}=\big[1+\int d\omega \frac{J_{m}(\omega )+(-1)^{j+1}J_{n}(\omega )}{(\omega -\varpi_{j}^{b})^2 }\big]^{-1}$. The QE entanglement in a state $\rho$ is measured by concurrence $\mathcal{C}_{2}=\max{\{0,\sqrt{\lambda_{1}}-\sqrt{\lambda_{2}}-\sqrt{\lambda_{3}}-\sqrt{\lambda_{4}}\}}$, where $\lambda_{i}$ are the eigenvalues of $\rho(\hat{\sigma}_{y}\otimes\hat{\sigma}_{y})\rho^{\ast}(\hat{\sigma}_{y}\otimes\hat{\sigma}_{y})$ in decreasing order \cite{PhysRevLett.80.2245}. From the bound-state analysis, we obtain
\begin{equation}\label{double-c}
\hspace{-.1cm}\mathcal{C}_2(\infty)=\left\{
\begin{array}{cc}
0, & M=0 \\
(K_{01}^{(1)})^2/2, & M=1 \\
|(K_{01}^{(1)})^{2}-(K_{01}^{(2)})^{2}+D(t)|/2, & M=2
\end{array},
\right.
\end{equation}
where $D(t)=2iK^{(1)}_{01}K^{(2)}_{01}\sin[(\varpi _{1}^{b}-\varpi _{2}^{b})t]$. It indicates that the features of the entanglement dynamics in the steady state is determined by the number of bound states, i.e., a stable or a persistently oscillating entanglement between the QEs would be generated once one or two bound states are formed. This is in sharp contrast to the previous Born-Markovian approximate results, where the entanglement approaches zero exclusively \cite{PhysRevB.84.235306,PhysRevLett.106.020501,PhysRevB.106.155409,acsphotonics,doi:10.1021/acsphotonics.7b00717,C9NR05083C,PhysRevB.96.075405}. These diverse signatures can be explained by our bound-state analysis (see Appendix \ref{Appen-dyna}). Revealing a mechanism to overcome the detrimental influence of the damping SPP in the absorptive medium on the generated entanglement between the QEs, this bound-state favored long-time feature endows our system with an ability to realize QI via the mediation role of the metal nanowire. Note that Eq. \eqref{double-c} reveals that the entanglement is lost at the degenerate point of the two bound states.

Choosing the metal as silver with $\varepsilon_\infty=5.7$, and $\hbar\gamma_p=0.1$ eV \cite{Scholl2012}, we show in Fig. \ref{Fig2}(a) the number of bound states via solving Eq. \eqref{eigen-solution}. It is seen that two bound states are formed in the small-$r_a$ regime. The entanglement evolution in Fig. \ref{Fig2}(b) reveals that, besides the severe oscillations in the transient dynamics characterizing the non-Markovian effect of rapidly energy exchange between the QEs caused by the near-field enhancement of the SPPs \cite{PhysRevB.82.115334,PhysRevB.95.075412}, rich entanglement features, e.g., complete decay, stable trapping, and Rabi-like oscillation, are exhibited in the steady state for different $r_a$. It indicates that the non-Markovian effect manifests its action on the QE not only in its transient dynamical process, but also in its steady state. The steady-state entanglement equals exactly to the ones predicted by Eq. \eqref{double-c} [see Fig. \ref{Fig2}(c)]. The regions where $\mathcal{C}_2(\infty)$ displays distinct features match well with the ones of different numbers of bound state formed in the energy spectrum [see Fig. \ref{Fig2}(d)]. Different from the asymptotic vanishing in the Born-Markovian approximation \cite{PhysRevLett.106.020501}, the result verifies our expectation that an efficient QI between the QEs with persistent entanglement is realized due to the formation of one or two bound states. Figures \ref{Fig2}(e)-\ref{Fig2}(h) show the results in different QE distance $d$. We see that a stable entanglement is generated for $d$ in support of the formation of a bound state for the relatively large QE-nanowire distance $r_a$ [see Figs. \ref{Fig2}(e) and \ref{Fig2}(f)]. With decreasing $r_a$, the QE-SPP coupling is so enhanced that two bound states are formed even when the QEs are separated in a distance as large as $0.3\lambda_0$. Then, a persistently oscillating entanglement can be established except for the accidental-degenerate points [see Figs. \ref{Fig2}(g) and \ref{Fig2}(h)]. With a fixed number of bound states, quantitative changes in values of the entanglement can be observed by changing the parameters, which indicates that the specific form of bound states determines the explicit values of entanglement. Efficiently avoiding the dissipation induced by the absorptive metal, the bound states are favorable for creating persistent entanglement between QEs. In contrast to the localized structure \cite{PhysRevResearch.1.023027}, the coupling of QEs with the transported SPPs along the  nanowire endows the generation of quantum entanglement between QEs in larger distance with more than one order of magnitude. Such bound-state favored behavior paves the way for applying SPPs in designing QI to scalable quantum network \cite{Tame2013,ZHOU20191}. Note that stronger confinement of the SPPs is accompanied with a higher metal absorption. We thus must make a balance to the strong QE-SPP coupling and the long propagation distance by choosing proper radius $R$ (see Appendix \ref{Appen-radiu}).

\begin{figure}[tbp]
\includegraphics[width=\columnwidth]{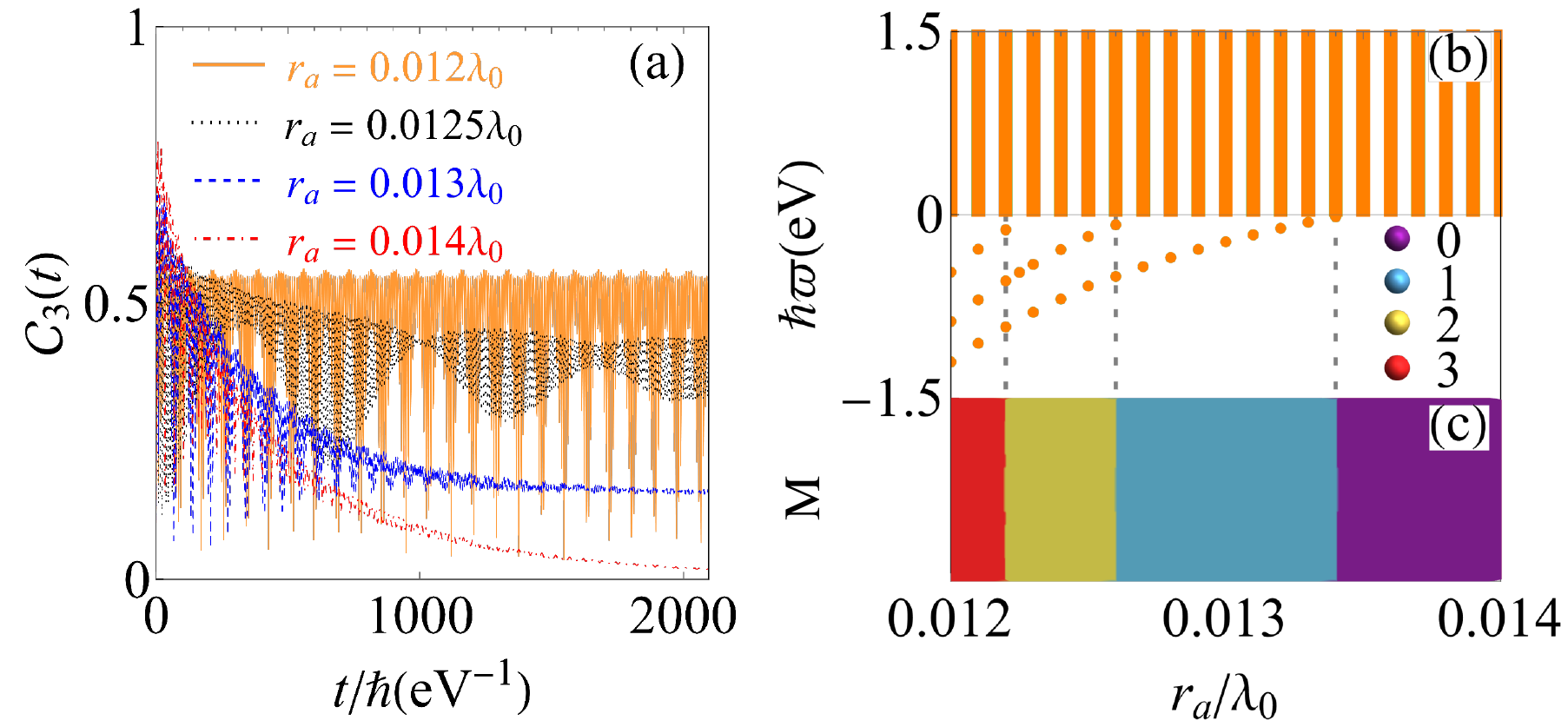}\\
\includegraphics[width=\columnwidth]{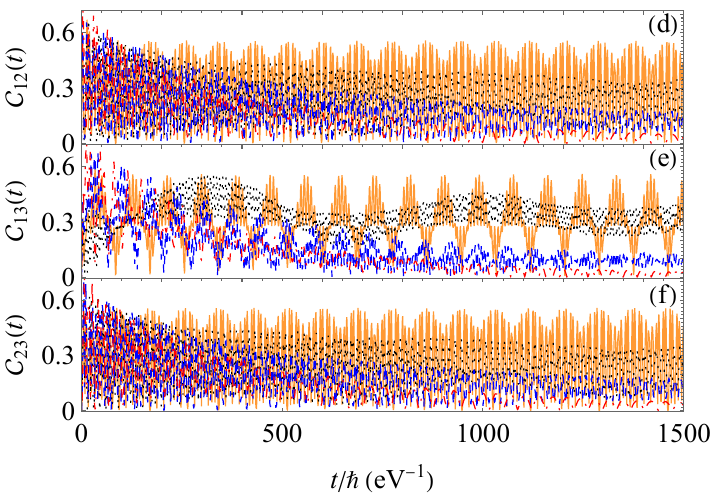}\\
\caption{(a) Evolution of $\mathcal{C}_{3}(t)$ in different $r_a$ by solving Eq. \eqref{evo-1}. (b) Energy spectrum and (c) numbers of the bound state as a function of $r_{a}$ by solving Eq. \eqref{eigen-solution}. (d-f) Evolution of $\mathcal{C}_{12}$, $\mathcal{C}_{13}$, and $\mathcal{C}_{23}$ in different $r_a$. Other parameters are the same as in Fig. \ref{Fig2}(b) except for $N=3$.}
\label{Fig3}
\end{figure}

\begin{figure}[tbp]
\centering
\includegraphics[width=.9\columnwidth]{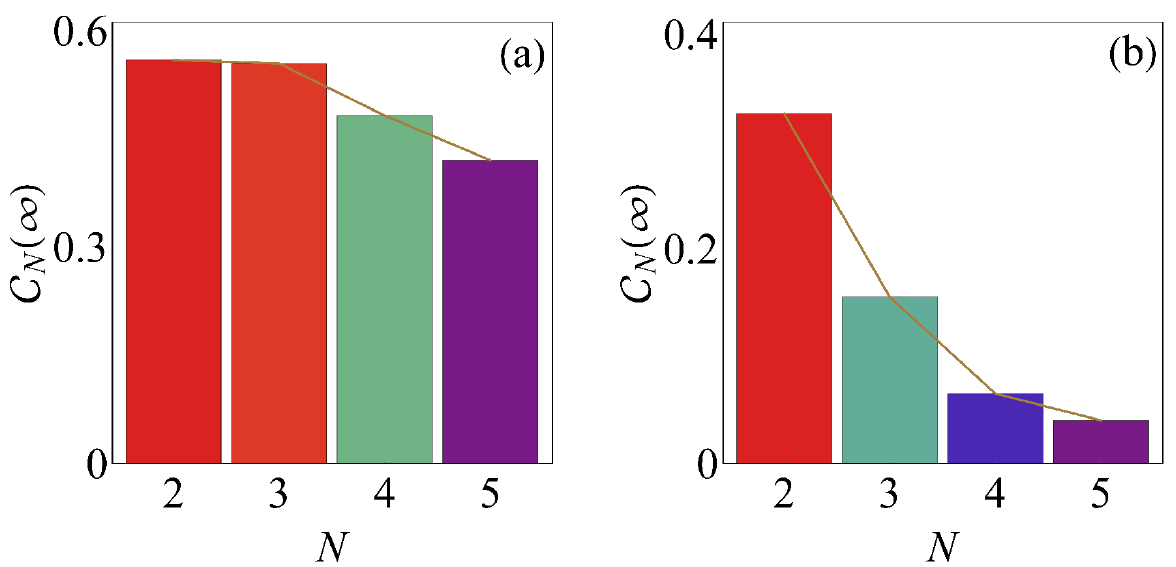}\\
\caption{$\mathcal{C}_{N}(\infty)$ in different $N$ when (a) $N$ and (b) one bound states are formed. Only the maxima of $\mathcal{C}_{N}(\infty)$ are plotted in (a). $r_a=0.012\lambda_0$ in (a) and $0.013\lambda_0$ in (b), others are the same as Fig. \ref{Fig2}(b).}
\label{Fig4}
\end{figure}

Our result can be scaled up to the cases with a large number of QEs. It is derived from Eqs. \eqref{eigen-solution} and \eqref{sol} the non-zero $\mathbf{Z}(t)$ when different numbers of bound states are formed. A lower bound of the $N$-qubit entanglement is $\mathcal{C}_{N}(\rho )\equiv [\frac{1}{N}\sum_{n=1}^{N}\sum_{k=1}^\mathcal{K}(\mathcal{C}_{k}^{n})^{2}]^{1\over2}$, where $\mathcal{C}_{k}^{n}=\max\{0,\sqrt{\lambda _{k,1}}-\sum_{q>1}\sqrt{\lambda_{k,q}}\}$, with $\lambda _{k,q}$ being the eigenvalues of $\rho L_{k}^{\bar{n}}\otimes \sigma_y^{n}\rho ^{\ast }L_{k}^{\bar{n}}\otimes \sigma_y^{n}$ in decreasing order with $q$. Here, $\sigma_y^n$ is the Pauli matrix acting on the $n$th qubit, $L_{k}^{\bar{n}}$ is the $k$th generator of the group SO($2^{N-1}$) acting on the other $N-1$ qubits, and $\mathcal{K}=2^{N-2}(2^{N-1}-1)$ is the number of the generators. $\mathcal{C}_{N}(\rho )$ reduces to the concurrence for $N=2$ \cite{Li_2009,RevModPhys.81.865}. Figure \ref{Fig3}(a) shows $\mathcal{C}_{3}(t)$ in different $r_a$ for $N=3$ under the condition $|\Psi(0)\rangle=|e_{1}g_{2}g_3\rangle $. Similar entanglement features as the $N=2$ case, e.g., complete decay, stable trapping, and Rabi-like oscillation but with more frequencies than the $N=2$ case, are present in the steady state. The regions of these features match well with the ones with zero, one, and more bound states formed in the energy spectrum [see Fig. \ref{Fig3}(b)]. At most three bound states can be formed for small $r_a$ [see Fig. \ref{Fig3}(c)]. The regions in the presence of the nonzero tripartite entanglement also support a non-vanishing bipartite entanglement. Figures \ref{Fig3}(d)-\ref{Fig3}(f) reveal that persistent entanglement between any pair of the QEs is also generated with the formation of bound states. We find in Fig. \ref{Fig4} that the bound-state favored stable entanglement is also valid for $N>3$. The above results reveal the scalability of our QI scheme.

\section{Discussion and conclusions}\label{con}
Our scheme is realizable in the state-of-the-art experiments. The QE may be a quantum dot or J aggregate, which can be excited by optical or electric driving \cite{Akimov2007,PhysRevLett.111.026804}. The nanowire was fabricated by synthetic and lithographic technologies \cite{PhysRevLett.106.226802}. The strong QE-SPP coupling is achievable by adjusting the QE positions via tuning the thickness of a spacer, the nanopositioning techniques, and the microfluidic flow control \cite{doi:10.1021/acs.chemrev.7b00441}. The controllable couplings between CdSe quantum dots or the J aggregate and SPPs and between multiple quantum dots and SPPs on a silver nanowire have been realized \cite{doi:10.1021/nl500838q,doi:10.1021/acs.nanolett.7b05448,Wu2021}. The bound state and its dominated role in the non-Markovian dynamics of one QE have been observed in circuit QED \cite{Liu2017} and cold-atom \cite{Krinner2018,Kwon2022} systems. This progress gives a strong support to the realization of our non-Markovian QI. Compared with the circuit QED \cite{Liu2017} and the cold-atom \cite{Krinner2018,Kwon2022} systems, the advantage of the SPPs is the strong QE-SPP coupling favored by the subdiffraction confinement and the room-temperature working condition. The optical-frequency feature of the SPPs makes them easily interact with the widely used solid-state qubits.

In summary, we have proposed a scheme of non-Markovian QI formed by a metal nanowire. A mechanism to generate a persistent entanglement among the distant QEs via the SPP supported by the metal nanowire is discovered. It is revealed that as long as one or more QE-SPP bound states are formed, the entanglement exhibits a finite-value stabilization or a persistent Rabi-like oscillation in the long-time limit, respectively. Overcoming the destructive impact of the metal absorption to the SPPs on the SPP-mediated QE entanglement, our result enriches our understanding on the light-matter interactions in lossy medium and supplies an instruction to apply SPP in designing scalable quantum devices.

\section{Acknowledgments}
This work is supported by the National Natural Science Foundation of China (Grants No. 12074106, No. 12275109, No. 12205128, No. 12247101, and No. 11834005) and China Postdoctoral Science Foundation (Grants No. BX20220138 and No. 2022M710063).

\appendix

\section{Macroscopic quantum electrodynamics}\label{Appen-quan}
The quantization of electromagnetic field in a dispersive, absorptive, nonmagnetic, and linear (homogeneous and isotropic) medium is phenomenologically realized by the method of macroscopic quantum electrodynamics. It is macroscopic because it treats the electromagnetic field as being governed by the macroscopic Maxwell equations and treats the medium in terms of permittivities and permeabilities, taking the charges and currents in the medium as continuous \cite{PhysRevA.53.1818}. The Maxwell equations in a dispersive and absorptive medium are
\begin{eqnarray}
{\pmb\nabla} \cdot \mathbf{B}(\mathbf{r})&=&0, ~{\pmb\nabla} \times \mathbf{E}(\mathbf{r})=-\mathbf{\dot{B}}(\mathbf{r}), \\
{\pmb\nabla} \cdot \mathbf{D}(\mathbf{r})&=&0,~ {\pmb\nabla} \times \mathbf{H}(\mathbf{r})=\mathbf{\dot{D}}(\mathbf{r}),
\end{eqnarray}
where $\mathbf{H}(\mathbf{r})=\mathbf{B}(\mathbf{r})/\mu _{0}$ and $\mathbf{D}=\varepsilon _{0}\mathbf{E}+\mathbf{P}$. Because the medium linearly responses to the electromagnetic field, the polarization $\mathbf{P}$ reads
\begin{equation}
\mathbf{P}(\mathbf{r},t)=\varepsilon _{0}\int_{0}^{\infty }\chi (\mathbf{r},\tau )\mathbf{E}(\mathbf{r},t-\tau )d\tau +\mathbf{P}_\text{N}(\mathbf{r},t),\label{dfds}
\end{equation}
where the convolution describes the material's polarization in response to the field and $\chi $ is the electric susceptibility of the medium. In Eq. \eqref{dfds}, we have introduced the noise polarization $\mathbf{P}_\text{N}$ to describe the medium absorption to the electromagnetic field. After making the Fourier transform, we obtain $\mathbf{D}(\mathbf{r},\omega )=\varepsilon _{0}\varepsilon (\omega )\mathbf{E}(\mathbf{r},\omega )+\mathbf{P}_\text{N}(\mathbf{r},\omega )$, where $\varepsilon (\omega )=1+\chi (\omega )$ is the relative permittivity. For the dispersive and absorptive medium, the permittivity $\varepsilon (\omega )$ is a complex function of frequency, i.e., $\varepsilon (\omega )=\varepsilon_\text{R} (\omega )+i\varepsilon_\text{I} (\omega )$. The real and imaginary parts, which are responsible for dispersion and absorption, respectively, are uniquely related to each other through the Kramers-Kronig relations as
\begin{eqnarray}
\varepsilon _\text{R}(\omega )-1&=&{\mathcal{P}\over\pi}\int d\omega'{\varepsilon_\text{I} (\omega )\over\omega'-\omega},\\
\varepsilon _\text{I}(\omega )&=&-{\mathcal{P}\over\pi}\int d\omega'{\varepsilon_\text{R} (\omega )-1\over\omega'-\omega},
\end{eqnarray}where $\mathcal{P}$ is the Cauchy's principal value. It is readily to derive the wave equation of the electric-field strength in the frequency domain as
\begin{equation}
[{\pmb\nabla}\times {\pmb\nabla} \times -\frac{\omega ^{2}}{c^{2}}\varepsilon (\omega )]\mathbf{E}(\mathbf{r},\omega )=i\omega\mu _{0}\mathbf{j}_\text{N}(\mathbf{r},\omega ),  \label{eq:3}
\end{equation}
where $\mathbf{j}_\text{N}(\mathbf{r},\omega )=-i\omega \mathbf{P}_\text{N}(\mathbf{r},\omega )$ is the noise current density. Equation \eqref{eq:3} can be solved by the Green's function method as
\begin{equation}
\mathbf{E}(\mathbf{r},\omega )=i\omega \mu_{0}\int d^{3}\mathbf{r}^{\prime }\mathbf{G}(\mathbf{r},\mathbf{r}^{\prime },\omega )\mathbf{j}_\text{N}(\mathbf{r}^{\prime },\omega )  \label{eq:1}
\end{equation}
with the Green's function $\mathbf{G}(\mathbf{r},\mathbf{r}^{\prime },\omega )$ determined by \cite{Rivera2020}
\begin{equation}
[{\pmb\nabla} \times {\pmb\nabla} \times -\frac{\omega ^{2}}{c^{2}}\varepsilon (\omega )]\mathbf{G}(\mathbf{r},\mathbf{r}^{\prime },\omega )=\delta (\mathbf{r}-\mathbf{r}^{\prime }).  \label{eq:2}
\end{equation}%
In macroscopic quantum electrodynamics, the transition from classical to quantum theory consists in the replacement of the classical noise current $\mathbf{j}_\text{N}(\mathbf{r},\omega)$ by the operator-valued field and the requirement of the noise current operator to satisfy the canonical commutation relation $[\hat{j}_{\text{N},i}(\mathbf{r},\omega ),\hat{j}_{\text{N},i}^{\dag }(\mathbf{r}^{\prime },\omega ^{\prime })]=\omega ^{2}\frac{\hbar \varepsilon _{0}}{\pi}\varepsilon _\text{I}(\omega )\delta _{ij}\delta (\mathbf{r}-\mathbf{r}^{\prime})\delta (\omega -\omega ^{\prime })$. This is equivalently achieved by setting $\hat{\mathbf{j}}_\text{N}(\mathbf{r},\omega )=\omega \sqrt{\frac{\hbar \varepsilon _{0}}{\pi }\varepsilon _\text{I}(\omega )}\hat{\mathbf{f}}(\mathbf{r},\omega )$, where $\hat{\mathbf{f}}(\mathbf{r},\omega )$ is the bosonic operator satisfying $[\hat{f}_{i}(\mathbf{r},\omega ),\hat{f}_{j}^{\dag }(\mathbf{r}^{\prime },\omega ^{\prime })]=\delta _{ij}\delta (\mathbf{r}-\mathbf{r}^{\prime})\delta (\omega -\omega ^{\prime })$\ and $[\hat{f}_{i}(\mathbf{r},\omega ),\hat{f}_{j}(\mathbf{r}^{\prime },\omega ^{\prime })]=[\hat{f}_{i}^{\dag }(\mathbf{r},\omega ),\hat{f}_{j}^{\dag }(\mathbf{r}^{\prime },\omega ^{\prime})]=0$. Therefore, we have the positive-frequency part of the quantized electric-field strength as
\begin{equation}
\hat{\mathbf{E}}(\mathbf{r},\omega )=i\sqrt{\frac{\hbar }{\pi \varepsilon_{0}}}\frac{\omega ^{2}}{c^{2}}\int d^{3}\mathbf{r}^{\prime }\sqrt{\varepsilon _\text{I}(\omega )}\mathbf{G}(\mathbf{r},\mathbf{r}^{\prime },\omega )\mathbf{\hat{f}}(\mathbf{r}^{\prime },\omega ).\label{eq:4}
\end{equation}

It is worth noting that when $\varepsilon_\text{I}(\omega)\rightarrow0$ in a certain frequency interval and hence $\hat{\bf j}_\text{N}({\bf r},\omega)\rightarrow0$, the electric field $\hat{\mathbf{E}}(\mathbf{r},\omega )$ in Eq. \eqref{eq:4} does not vanish. A careful inspection of the integral reveals that $\hat{\mathbf{E}}(\mathbf{r},\omega )$ tends to the familiar representation of the source-free field through mode decomposition. In particular, in the absence of dielectric matter, i.e., $\varepsilon(\omega)\rightarrow 1$, it reduces to the source-free field in free space \cite{PhysRevA.53.1818}. To prove this, we take a linearly polarized electric field propagating in the $x$ direction as an example, which effectively reduces the system to one spatial dimension, i.e., $\hat{\mathbf{E}}(\mathbf{r},\omega )\rightarrow E_y(x,\omega)\equiv E(x,\omega)$, $\hat{\bf f}({\bf r},\omega)\rightarrow \hat{f}_y(x,\omega)\equiv \hat{f}(x,\omega)$, and $\mathbf{G}(\mathbf{r},\mathbf{r}^{\prime },\omega )\rightarrow G(x,x',\omega)$. Then, the wave equation \eqref{eq:2} takes the form
\begin{equation}
{\partial^2G(x,x',\omega)\over\partial x^2}+{\omega^2\varepsilon(\omega)\over c^2}G(x,x',\omega)=-\delta(x-x').
\end{equation}
Using Cauchy's residue theorem, its solution is calculated as
\begin{eqnarray}
G(x,x',\omega)=\int{dk\over 2\pi}{c^2 e^{ik(x-x')}\over c^2k^2-\omega^2\varepsilon(\omega)}={ie^{i{\omega n(\omega)\over c}|x-x'|}\over 2c^{-1}\omega n(\omega)},~\label{sgrf}
\end{eqnarray}where $n(\omega)=\sqrt{\varepsilon(\omega)}\equiv\beta(\omega)+i\gamma(\omega)$ is the complex refractive index. Substituting Eq. \eqref{sgrf} into Eq. \eqref{eq:4}, we obtain
\begin{eqnarray}
E(x,\omega)&=&i\sqrt{\hbar\over 4\pi c\omega \varepsilon_0\beta(\omega)\mathcal{A}}{\beta(\omega)\omega\over n(\omega)}\Big[e^{i\omega\beta(\omega) x\over c}\hat{a}_+(x,\omega)\nonumber\\&&+e^{-i\omega\beta(\omega) x\over c}\hat{a}_-(x,\omega)\Big],\label{smelct}
\end{eqnarray}
where $\mathcal{A}$ is the normalization area perpendicular to the $x$ direction and
\begin{eqnarray}\hat{a}_\pm(x,\omega)&=&i\sqrt{2\gamma(\omega)\over c\omega^{-1}}{e^{\mp\gamma(\omega) x\over c\omega^{-1}}}\nonumber\\&&\times\int_{-\infty}^{\pm x}e^{-i n(\omega) x'\over c\omega^{-1}}\hat{f}(\pm x',\omega)dx'.\label{smanno}
\end{eqnarray}
The subscripts $+$ and $-$ denote propagation to the right and left, respectively. It can be derived that
\begin{equation}
[\hat{a}_\pm(x,\omega),\hat{a}^\dag_\pm(x',\omega])=e^{-\gamma(\omega)\omega |x-x|/c}\delta(\omega-\omega').\label{smcmt}
\end{equation}
Now, assuming that the absorption of the medium to the field in a chosen frequency interval $\Delta\omega$ is so small that the condition $\gamma(\omega)\ll\beta(\omega)$ is satisfied. Equations \eqref{smanno} and \eqref{smcmt} indicate that the operators $\hat{a}_\pm(x,\omega)$ becomes independent of $x$ when $\gamma(x)\omega |x-x'|/c\rightarrow 0$. Then, Eq. \eqref{smelct} reduces, in the chosen frequency interval $\Delta\omega$, to the familiar result
\begin{eqnarray}
E(x,\omega)&=&i\omega\sqrt{\hbar\over 4\pi c\omega \varepsilon_0\beta(\omega)\mathcal{A}}\Big[e^{i\omega\beta(\omega) x\over c}\hat{a}_+(\omega)\nonumber\\&&+e^{-i\omega\beta(\omega) x\over c}\hat{a}_-(\omega)\Big],
\end{eqnarray}
where the associated photon operators $\hat{a}_\pm(\omega)$ and $\hat{a}^\dag_\pm(\omega)$ satisfy the commutation relation $[\hat{a}_\pm(\omega),\hat{a}^\dag_\pm(\omega')]=\delta(\omega-\omega')$. Further when the dielectric matter is absent, i.e., $\beta(\omega)=1$, the electric-field strength becomes the quantized form in the free space
\begin{eqnarray}
E(x,\omega)=i\omega\sqrt{\hbar\over 4\pi c\omega \varepsilon_0\mathcal{A}}\Big[e^{i\omega x\over c}\hat{a}_+(\omega)+e^{-i\omega x\over c}\hat{a}_-(\omega)\Big],~~~
\end{eqnarray}where $\omega=c k$. It is obvious that the field operator becomes the familiar free-space form in the limit $\varepsilon_\text{I}(\omega)\rightarrow0$ \cite{PhysRevA.53.1818}.

\section{Green's tensor}\label{Appen-Green}
We here provide the analytical form of the Green's tensor of the electromagnetic field in the interface of a metallic nanowire and the free space. Consider an infinitely long cylindrical nanowire with a complex dielectric function $\varepsilon _{m}(\omega )$ and a radius $R$. The Green's tensor at the observation point ${\bf r}$ triggered by a source point at $\mathbf{r}^{\prime }$ outside the cylinder is \cite{PhysRevB.82.075427}
\begin{equation}
\mathbf{G}(\mathbf{r},\mathbf{r}^{\prime },\omega )=\left\{\begin{array}{cc}
\mathbf{G}^{0}(\mathbf{r},\mathbf{r}^{\prime },\omega )+\mathbf{G}^{R}(\mathbf{r},\mathbf{r}^{\prime },\omega ), & r>R \\
\mathbf{G}^{T}(\mathbf{r},\mathbf{r}^{\prime },\omega ), & r<R
\end{array}
\right.
\end{equation}
where $\mathbf{G}^{0}(\mathbf{r},\mathbf{r}^{\prime},\omega)$, $\mathbf{G}^{R}(\mathbf{r},\mathbf{r}^{\prime},\omega)$, and $\mathbf{G}^{T}(\mathbf{r},\mathbf{r}^{\prime},\omega)$ are the Green's tensors contributed by the free-space, the reflected, and the transmitted fields, respectively. To fulfill the boundary conditions, we expand the Green's tensor in cylindrical harmonics. Then, we have
\begin{widetext}
\begin{eqnarray}
\mathbf{G}^{0}(\mathbf{r},\mathbf{r}^{\prime },\omega ) &=&-\frac{\mathbf{%
\hat{r}\hat{r}}\delta (\mathbf{r}-\mathbf{r}^{\prime })}{k_{0}^{2}}+\frac{i}{%
8\pi }\int_{-\infty }^{\infty }dk_{z}\sum_{n=0}^{\infty }\frac{2-\delta
_{n,0}}{k_{r_{0}}^{2}} \\
&&\times \left\{
\begin{array}{cc}
\lbrack \mathbf{M}_{_{o}^{e}n}^{(1)}(k_{r_{0}},k_{z},\mathbf{r})\mathbf{M}%
_{_{o}^{e}n}(k_{r_{0}},-k_{z},\mathbf{r}^{\prime })+\mathbf{N}%
_{_{o}^{e}n}^{(1)}(k_{r_{0}},k_{z},\mathbf{r})\mathbf{N}%
_{_{o}^{e}n}(k_{r_{0}},-k_{z},\mathbf{r}^{\prime })], & \mathbf{r>r}^{\prime
}, \\
\lbrack \mathbf{M}_{_{o}^{e}n}(k_{r_{0}},k_{z},\mathbf{r})\mathbf{M}%
_{_{o}^{e}n}^{(1)}(k_{r_{0}},-k_{z},\mathbf{r}^{\prime })+\mathbf{N}%
_{_{o}^{e}n}(k_{r_{0}},k_{z},\mathbf{r})\mathbf{N}%
_{_{o}^{e}n}^{(1)}(k_{r_{0}},-k_{z},\mathbf{r}^{\prime })], & \mathbf{r<r}%
^{\prime },%
\end{array}%
\right.   \nonumber \\
\mathbf{G}^{R}(\mathbf{r},\mathbf{r}^{\prime },\omega ) &=&\frac{i}{8\pi }%
\int_{-\infty }^{\infty }dk_{z}\sum_{n=0}^{\infty }\frac{2-\delta _{n0}}{%
k_{r_{0}}^{2}}\times \{[A_{R_{o}^{e}}\mathbf{M}%
_{_{o}^{e}n}^{(1)}(k_{r_{0}},k_{z},\mathbf{r})+B_{R_{e}^{o}}\mathbf{N}%
_{_{e}^{o}n}^{(1)}(k_{r_{0}},k_{z},\mathbf{r})]\mathbf{M}%
_{_{o}^{e}n}^{(1)}(k_{r_{0}},-k_{z},\mathbf{r}^{\prime }) \\
&&+[C_{R_{o}^{e}}\mathbf{N}_{_{o}^{e}n}^{(1)}(k_{r_{0}},k_{z},\mathbf{r}%
)+D_{R_{e}^{o}}\mathbf{M}_{_{e}^{o}n}^{(1)}(k_{r_{0}},k_{z},\mathbf{r})]%
\mathbf{N}_{_{o}^{e}n}^{(1)}(k_{r_{0}},-k_{z},\mathbf{r}^{\prime })\},
\nonumber \\
\mathbf{G}^{T}(\mathbf{r},\mathbf{r}^{\prime },\omega ) &=&\frac{i}{8\pi }%
\int_{-\infty }^{\infty }dk_{z}\sum_{n=0}^{\infty }\frac{2-\delta _{n0}}{%
k_{r_{0}}^{2}}\times \{[A_{T_{o}^{e}}\mathbf{M}_{_{o}^{e}n}(k_{r_{1}},k_{z},
\mathbf{r})+B_{T_{e}^{o}}\mathbf{N}_{_{e}^{o}n}(k_{r_{1}},k_{z},\mathbf{r})]
\mathbf{M}_{_{o}^{e}n}^{(1)}(k_{r_{0}},-k_{z},\mathbf{r}^{\prime }) \\
&&+[C_{T_{o}^{e}}\mathbf{N}_{_{o}^{e}n}(k_{r_{1}},k_{z},\mathbf{r}%
)+D_{T_{e}^{o}}\mathbf{M}_{_{e}^{o}n}(k_{r_{1}},k_{z},\mathbf{r})]\mathbf{N}
_{_{o}^{e}n}^{(1)}(k_{r_{0}},-k_{z},\mathbf{r}^{\prime })\},  \nonumber
\end{eqnarray}
where $A_{R_{o}^{e}}$, $B_{R_{e}^{o}},C_{R_{o}^{e}}$, $D_{R_{e}^{o}}$ and $A_{T_{o}^{e}}$, $B_{T_{e}^{o}},C_{T_{o}^{e}}$, $D_{T_{e}^{o}}$ are the reflection and transmission coefficients, the subscript $e$ and $o$ denote even and odd, $\mathbf{M}_{_{o}^{e}n}(k_{r_{0/1}},k_{z},\mathbf{r})={\pmb\nabla} \times \lbrack J_{n}(k_{r_{0/1}}r)(_{\sin n\phi }^{\cos n\phi })e^{ik_{z}z}\mathbf{z}]$ and $\mathbf{N}_{_{o}^{e}n}(k_{r_{0/1}},k_{z},\mathbf{r})=\frac{1}{k_{0/1}}{\pmb \nabla} \times \mathbf{M}_{_{o}^{e}n}(k_{z},\mathbf{r})$ are the cylindrical harmonic vector wave functions, respectively. The tensor product is defined as $\mathbf{M}_{_{e}^{o}n}\mathbf{N}_{_{o}^{e}n}^{(1)}=\mathbf{M}_{o,n}\mathbf{N}_{e,n}^{(1)}+\mathbf{M}_{e,n}\mathbf{N}_{o,n}^{(1)}$. Here, $\mathbf{r}=(r,\phi,z)$, $\mathbf{k}=(k,k_{\phi},k_z)$, and $k_{r_{0/1}}=\sqrt{k_{0/1}^{2}-k_{z}^{2}}$ with $0$ and $1$ denoting the wave vectors outside and inside the cylinder. When ``(1)" appears at the superscript of $\mathbf{M}$ or $\mathbf{N}$, the Bessel function $J_{n}(x)$ has to be replaced by the Hankel function of the first kind $H_{n}^{(1)}(x)$. According to the boundary conditions $\mathbf{\hat{r}}\times \lbrack \mathbf{G}({\bf r},{\bf r}^{\prime })_{{r}={R}^{-}}-\mathbf{G}({\bf r},{\bf r}^{\prime })_{{ r}={R}^{+}}]=0$ and $\mathbf{\hat{r}}\times {\pmb \nabla} \times \lbrack \mathbf{G}({\bf r},{\bf r}^{\prime})_{{r}={R}^{-}}-\mathbf{G}({\bf r},{\bf r}^{\prime })_{{r}={R}^{+}}]=0$, the reflection and transmission coefficients can be determined.

The coupling between the quantum emitter (QE) and the surface plasmon polariton (SPP) is strongest when the dipole moment of the QE orients towards the radial direction \cite{PhysRevB.84.235306}. Then only the $rr$ component of the Green's tensor contributes to the interactions. We obtain
\begin{equation}
\mathbf{G}_{rr}(\mathbf{r},\mathbf{r}^{\prime },\omega )=-\frac{\delta (\mathbf{r}-\mathbf{r}^{\prime })}{k_{0}^{2}}+\frac{i}{8\pi }\int_{-\infty
}^{\infty }dk_{z}\sum_{n=0}^{\infty }(2-\delta _{n,0})e^{ik_{z}(z-z^{\prime
})}\xi _{n}(k_{r_{0}}r),
\end{equation}
with
\begin{equation}
\xi _{n}(x)=\frac{n^{2}H_{n}^{(1)}(x)}{x^{2}}[J_{n}(x)+A_{R^{e}}H_{n}^{(1)}(x)]+\frac{k_{z}^{2}\partial _{r}H_{n}^{(1)}(x)}{k_{0}^{2}k_{r_{0}}^{2}}\partial _{r}[J_{n}(x)+C_{R^{e}}H_{n}^{(1)}(x)]+\frac{2inB_{R_{e}}k_{z}}{k_{0}k_{r_{0}}x}H_{n}^{(1)}(x)\partial
_{r}H_{n}^{(1)}(x)
\end{equation}
where the contributions of the free-space, the reflected, and the transmitted fields have been incorporated. The spectral density characterizing the couplings between QEs induced by the SPP is thus obtained as
\begin{equation}
J_{lj}(\omega )=\frac{3\gamma _{0}c\omega ^{2}}{8\pi \omega _{0}^{3}}\text{Im%
}[\int_{-\infty }^{\infty }dk_{z}\sum_{n=0}^{\infty }(2-\delta
_{n,0})e^{ik_{z}(z_l-z_j)}\xi _{n}(k_{r_{0}}r)],\label{appsyn}
\end{equation}
where $\gamma _{0}=\omega _{0}^{3}\mu ^{2}/3\pi \hbar \varepsilon _{0}c^{3}$ is the decay rate of the QEs in the free space. In Ref. \cite{PhysRevB.82.075427}, the fundamental $n=0$ mode of the SPPs in nanowire has been studied to create quantum gate between two $\Lambda-$type QEs under the Born-Markovian approximation. Here, we provide a rigorous study on the dynamics of an array of QEs coupling with the nanowire with all orders of the SPPs taken into account and investigate the positive roles of non-Markovian effect in designing QI.
\end{widetext}

\section{Energy spectrum}\label{Appen-eigenenergy}
The eigenenergies of the hybrid QE-SPP system can be derived as follows. In the single-excitation subspace, the eigenstate of the total QE-SPP system is expanded as
\begin{equation}
|\Phi\rangle =[\sum_{l}x_{l}\hat{\sigma}_{l}^{\dag}+\int d^{3}\mathbf{r}\int
d\omega y_{\mathbf{r},\omega}\mathbf{\hat{f}}^{\dag }(\mathbf{r},\omega )]|G;\{0_{\omega }\}\rangle.\label{eddgst}\end{equation} From the stationary Schr\"{o}dinger equation $\hat{H}|\Phi\rangle =E|\Phi\rangle $, with $E$ being the eigenenergy of the total system, we have
\begin{eqnarray}
Ex_{l} &=&\hbar \omega _{0}x_{l}-\frac{ic^{-2}\omega ^{2}}{\sqrt{\pi \varepsilon _{0}/\hbar }} \int d^{3}\mathbf{r}\int d\omega\sqrt{\text{Im}[\varepsilon _{m}(\omega )]}\nonumber\\
&&\times{\pmb\mu} _{l}\cdot{\bf G}(\mathbf{r}_{l},\mathbf{r},\omega )y_{\mathbf{r},\omega }, \label{B1}\\
Ey_{\mathbf{r},\omega } &=&\hbar \omega_0 y_{\mathbf{r},\omega }+\frac{i c^{-2}\omega ^{2}}{\sqrt{\pi \varepsilon _{0}/\hbar }}\sqrt{\text{Im}[\varepsilon _{m}(\omega )]}\nonumber\\
&&\times\sum_{l}{\pmb\mu} _{l}^{\ast }\cdot{\bf G}^{\ast }(\mathbf{r}_{l},\mathbf{r},\omega )x_{l}.\label{B2}
\end{eqnarray}
Substituting $y_{\mathbf{r},\omega }$ solved from Eq. \eqref{B2} into Eq. \eqref{B1}, we have
\begin{equation}\label{Bo}
\lbrack E-\hbar \omega _{0}-\hbar ^{2}\int d\omega \frac{\mathbf{J}(\omega )
}{E-\hbar \omega }]\mathbf{x}=0,
\end{equation}%
where $\mathbf{x}=\left( \begin{array}{ccc}  x_1 & \cdots & x_N \\  \end{array}\right)^\text{T}$. In the derivation, the relation as $\int d^{3}s\frac{\omega ^{2}}{c^{2}}\text{Im}[\varepsilon_{m}(\omega )]{\bf G}({\bf r},{\bf s},\omega ){\bf G}^{\ast }({\bf r}^{\prime },{\bf s},\omega )=\text{Im}[{\bf G}({\bf r},{\bf r}^{\prime},\omega )]$ and the definition of the spectral density have been used. The equations have nontrivial solution if and only if the determinant of the coefficient matrix is zero. Therefore, we obtain the eigen-equations
\begin{equation}\label{eigenenergy}
y_l({E\over\hbar})\equiv \omega _{0}+\int d\omega \frac{D_{l}(\omega )}{{E\over\hbar}-\omega }={E\over\hbar},~l=1,\cdot\cdot\cdot,N,
\end{equation}
where the Jordan decomposition $\mathbf{J}(\omega )=\mathbf{VD}(\omega )\mathbf{V}^{-1}$, with $\mathbf{D}(\omega )=$diag$[D_{1}(\omega ),\cdots,D_{N}(\omega )] $ being the Jordan canonical form of $\mathbf{J}(\omega )$ under the similarity transform $\mathbf{V}$, is used. It is evident that Eq. \eqref{eigenenergy} has the same form as Eq. (3) in the main text determining the dynamics of the QEs. This indicates that the dynamics of QEs is intrinsically determined by the energy-spectrum character of the total QE-SPP system. Because $y_l({E\over\hbar})$ is a decreasing function in the region of $E<0$, Eq. \eqref{eigenenergy} has discrete root $E^b$ provided $y_l(0)<0$. The hybrid eigenstate corresponding to such a discrete $E^b$ is called a bound state, which is the eigenstate of the total QE-SPP system with the corresponding eigenenergy falling in the band-gap regime. In the region of $E>0$, $y_l({E\over\hbar})$ is ill defined and thus Eq. \eqref{eigenenergy} has an infinite number of roots, which form a continuous energy band.

The excited-state amplitude $\mathbf{x}$ can be calculated by substituting Eqs. \eqref{B1} and \eqref{B2} into the normalization condition $\sum_{l}|x_{l}|^{2}+\int d^{3}\mathbf{r}\int d\omega|y_{\mathbf{r}, \omega }|^{2}=1$. In the case of $N=2$, there are at most two bound states formed. The Jordan canonical form of the spectral density $\mathbf{J}(\omega )=\left(
\begin{array}{cc}
J_{0}(\omega ) & J_{1}(\omega ) \\
J_{1}(\omega ) & J_{0}(\omega )%
\end{array}%
\right)
$ reads
\begin{equation}
{\bf D}(\omega)=\text{diag}\left[
                             \begin{array}{cc}
                               J_0(\omega)+J_1(\omega), & J_0(\omega)-J_1(\omega) \\
                             \end{array}
                           \right].
\end{equation} When one bound state with eigenenergy $E^b$ is formed under $y_1({E\over\hbar})<0$, we have $x^b_{1}=x^b_{2}$ and
\begin{eqnarray}
|x_1^b|^2={1\over 2}\Big[1+\int d\omega{J_0(\omega)-J_1(\omega)\over (E^b/\hbar-\omega)^2}\Big]^{-1}.
\end{eqnarray}When one bound state with eigenenergy $E^b$ is formed under $y_2({E\over\hbar})<0$, we have $x^b_{1}=-x^b_{2}$ and
\begin{eqnarray}
|x_1^b|^2={1\over 2}\Big[1+\int d\omega{J_0(\omega)+J_1(\omega)\over (E^b/\hbar-\omega)^2}\Big]^{-1}.
\end{eqnarray}
When two bound states with eigenenergies $E^b_j$ ($j=1,2$) are formed, we have $M=2$, $x^b_{j,1}=(-1)^{j+1}x^b_{j,2}$, and
\begin{equation}\label{S13}
|x^b_{j,1}|^{2}=\frac{1}{2}\Big[1+\int d\omega \frac{J_{0}(\omega)+(-1)^{j+1} J_{1}(\omega)}{(E^{b}_{j}/\hbar- \omega )^{2}}\Big]^{-1}.
\end{equation}

\section{Steady-state solution}\label{Appen-solution}
Without an analytical solution, Eq. (2) in the main text can only be numerically solved. However, we can analytically derive its asymptotical form in the long-time condition. A Laplace transform $\mathbf{\bar{c}}(s)=\int_{0}^{\infty }e^{-st}\mathbf{c}(t)dt$ recasts Eq. (2) into $\mathbf{\bar{c}}(s)=\big[ s+i\omega _{0}+\int_{0}^{\infty} d\omega \frac{\mathbf{J}(\omega )}{s+i\omega }\big] ^{-1}\mathbf{c}(0)$.
\subsection{System with $N=2$}
Remembering the Jordan canonical form ${\bf D}(\omega)=\text{diag}\left[
                             \begin{array}{cc}
                               J_0(\omega)+J_1(\omega), & J_0(\omega)-J_1(\omega) \\
                             \end{array}
                           \right]$ and under the initial state $\mathbf{c}(0)=(1,0)^\text{T}$, $\mathbf{\bar{c}}(s)$ equals to
\begin{equation}
\mathbf{\bar{c}}(s)=\frac{1}{2}\sum_{j=1}^{2}\frac{1}{\mathcal{K}_{01}^{(j)}(s)}\left(
\begin{array}{c}
1 \\
(-1)^{j+1}%
\end{array}%
\right),
\end{equation}%
with $\mathcal{K}_{mn}^{(j)}(s)=s+i\omega _{0}+\int d\omega \frac{J_{m}(\omega )+(-1)^{j+1}J_{n}(\omega )}{s+i\omega}$. Its inverse Laplace transform reads $\mathbf{c}(t)=-\frac{1}{2\pi }\int_{i\sigma +\infty }^{i\sigma -\infty }\mathbf{\bar{c}}(-i\varpi )e^{-i\varpi t}d\varpi $, which can be done by finding the poles of $\mathcal{K}_{01}^{j}(\varpi )=0$, i.e.,
\begin{equation}
y_j(\varpi)\equiv\omega _{0}+ \int d\omega \frac{D_{j}(\omega )}{\varpi-\omega }=\varpi, \label{smpol}
\end{equation}
which matches exactly with Eq. \eqref{eigenenergy} under the relation $E=\hbar\varpi$.
 According to the Cauchy's theorem, we obtain the general solution of Eq. (4) as
\begin{equation}
\mathbf{c}(t)=\mathbf{Z}(t)+\int_{0}^{\infty}\frac{d\varpi }{2\pi }[\mathbf{\bar{c}}(0^{+}-i\varpi)-\mathbf{\bar{c}}(0^{-}-i\varpi)]e^{-i\varpi t},\label{smsol}
\end{equation}
where the first and the second terms are contributed from the isolated poles $\varpi_j^b$ and the branch cut of Eq. \eqref{smpol}, respectively. We call the eigenstates corresponding to the isolated poles $E^b_j=\hbar\varpi_j^b$ the bound states. In the long-time limit, the second term tends to zero due to the out-of-phase interference of the poles along the branch cut. Thus, only the contributions of the bound states, i.e., $\lim_{t\rightarrow \infty }\mathbf{c}(t)=\mathbf{Z}(t)$, with $\mathbf{Z}(t)=-i\sum_{j=1}^{M}$Res$[\mathbf{\bar{c}}(-i\varpi_{j}^{b})e^{-i\varpi _{j}^{b}t}]$ being the residues of the isolated poles and $M$ being the number of formed bound states, survive. Using the L'Hospital rule, the residues can be obtained. When one bound
state with eigenenergy $\varpi ^{b}$\ is formed under $y_{j}(\varpi ^{b})<0,$%
\ we have
\begin{equation}
\mathbf{Z}(t)=\frac{1}{2}K_{01}^{(j)}e^{-i\varpi ^{b}t}\left(
\begin{array}{c}
1 \\
(-1)^{j+1}%
\end{array}%
\right) ,\label{smsing}
\end{equation}%
with $K_{01}^{(j)}=[\partial _{\varpi }\mathcal{K}_{01}^{(j)}(\varpi
)|_{\varpi =\varpi ^{b}}]^{-1}.$\ When two bound states with eigenenergies $%
\varpi _{j}^{b}$\ are formed, we have $M=2$\ and\  \
\begin{equation}
\mathbf{Z}(t)=\frac{1}{2}\sum_{j=1}^{M}K_{01}^{(j)}e^{-i\varpi
_{j}^{b}t}\left(
\begin{array}{c}
1 \\
(-1)^{j+1}%
\end{array}%
\right) ,\label{smtwo}
\end{equation}%
with $K_{01}^{(j)}=[\partial _{\varpi }\mathcal{K}_{01}^{(j)}(\varpi
)|_{\varpi =\varpi _{j}^{b}}]^{-1}.$\ From the definition, it is easy to
obtain $\partial _{\varpi }\mathcal{K}_{01}^{(j)}(\varpi )=1+\int d\omega
\frac{J_{m}(\omega )+(-1)^{j+1}J_{n}(\omega )}{(\omega -\varpi )^{2}}.$
According to the form of $x_{j,1}^{b}$ obtained in Eq. \eqref{S13}, we have $K_{01}^{(j)}=2|x_{j,1}^{b}|^{2}$.

\subsection{System with $N=3$}
The Jordan canonical form of the spectral density $\mathbf{J}(\omega )=\left(
\begin{array}{ccc}
J_{0}(\omega ) & J_{1}(\omega ) & J_{2}(\omega ) \\
J_{1}(\omega ) & J_{0}(\omega ) & J_{1}(\omega ) \\
J_{2}(\omega ) & J_{1}(\omega ) & J_{0}(\omega )%
\end{array}%
\right) $ reads
\begin{eqnarray}
\mathbf{D}(\omega ) &=&\text{diag}[J_{0}(\omega )-J_{2}(\omega ),  \notag \\
&&\frac{1}{2}[2J_{0}(\omega )+J_{2}(\omega )-\sqrt{8J_{1}^{2}(\omega
)+J_{2}^{2}(\omega )}],  \notag \\
&&\frac{1}{2}[2J_{0}(\omega )+J_{2}(\omega )+\sqrt{8J_{1}^{2}(\omega
)+J_{2}^{2}(\omega )]}]. ~~~~~~
\end{eqnarray}
Under the initial condition $\mathbf{c}(0)=(1,0,0)^\text{T}$, we have
\begin{equation}
\mathbf{\bar{c}}(s)=\frac{1}{2}\left(
\begin{array}{c}
\mathcal{K}_{0}(s)Y^{-1}(s)-[\mathcal{K}_{02}^{(2)}(s)]^{-1} \\
2Y^{-1}(s)[s+i\omega _{0}-\mathcal{K}_{1}(s)] \\
\mathcal{K}_{0}(s)Y^{-1}(s)+[\mathcal{K}_{02}^{(2)}(s)]^{-1}%
\end{array}%
\right) ,
\end{equation}
where $Y(s)=\mathcal{K}_{0}(s)\mathcal{K}_{02}^{(1)}(s)-2[\mathcal{K}_{1}(s)-s-i\omega _{0}]^{2}$ and $\mathcal{K}_{m}(s)=s+i\omega _{0}+\int d\omega \frac{J_{m}(\omega )}{s+i\omega }$. ${\bf c}(t)$ is obtainable by the inverse Laplace transform to $\mathbf{\bar{c}}(s)$, which needs find the poles of $Y(s)$ and $\mathcal{K}_{02}^{(2)}(s)$. Being similarly to the $N=2$ case, the contributions of the branch cuts of $Y(s)=0$ and $\mathcal{K}_{02}^{(2)}(s)=0$ tend to zero and only the ones of their isolated poles survive in the long-time condition. The contributions of these isolated poles can be derived via the residue theorem. After a straightforward calculation, we obtain the asymptotical solutions $\lim_{t\rightarrow \infty }\mathbf{c}(t)=\mathbf{Z}(t)=[Z_{1}(t),Z_{2}(t),Z_{3}(t)]^\text{T}$, where
\begin{eqnarray}
Z_{^1_3
}(t) &=&\frac{-i}{2}\{\sum_{j=1}^{M^{\prime }}\mathcal{K}_{0}(-i\varpi
_{j}^{b})[\partial _{\varpi }Y(-i\varpi _{j}^{b})]^{-1}e^{-i\varpi
_{j}^{b}t}\nonumber\\
&&\mp[\partial _{\varpi }\mathcal{K}_{02}^{(2)}(-i\varpi ^{b})]^{-1}e^{-i\varpi
^{b}t}\}, \\
Z_{2}(t) &=&-i\sum_{j=1}^{M^{\prime }}[-i\varpi
_{j}^{b}+i\omega _{0}-\mathcal{K}_{1}(-i\varpi _{j}^{b})]\nonumber\\
&&\times[\partial _{\varpi }Y(-i\varpi _{j}^{b})]^{-1}e^{-i\varpi
_{j}^{b}t}.
\end{eqnarray}
Here $\varpi _{j}^{b}$\ and $\varpi ^{b}$ are the isolated poles of $Y(-i\varpi)=0$ and $\mathcal{K}_{02}^{(2)}(-i\varpi)=0$, respectively.

\section{Dynamical effects of bound states}\label{Appen-dyna}
Consider that $M$ bound states are formed. According to the completeness of the eigenstates $\hat{H}|\Phi\rangle =E|\Phi\rangle $, we can expand the time evolved state as
\begin{equation}\label{S24}
|\Psi(t)\rangle=\sum_{j=1}^M c^{b}_{j} e^{-iE^b_j t\over\hbar}|\Phi^b_j\rangle+\sum_{\alpha\in \text{Band}}c_\alpha e^{-i E_{\alpha} t\over\hbar}|\Phi_\alpha\rangle,
\end{equation}
where $c^{b}_j=\langle \Phi_j^b|\Psi(0)\rangle$ and $c_\alpha=\langle \Phi_\alpha|\Psi(0)\rangle$. The first term is contributed from the bound states and the second one is from the continuous-band eigenstates.
The excited-state population of each QE, i.e., $P_l(t)=\langle\Psi(t)|\hat{\sigma}_l^\dag\hat{\sigma}_l|\Psi(t)\rangle$, is calculated to be
\begin{eqnarray}
P_l(t)&=&\sum_{j,j'=1}^M c^{b*}_{j'}c^{b}_je^{-i(E_j^b-E_{j'}^b)t\over\hbar} x^{b*}_{j',l}x^b_{j,l}\nonumber\\
&&+\sum_{\alpha,\alpha'\in \text{Band}} c^{*}_{\alpha'} c_\alpha e^{-i(E_\alpha-E_{\alpha'})t\over\hbar}  x^{*}_{\alpha',l}x_{\alpha,l}\nonumber\\
&&+\sum_{j=1}^M\sum_{\alpha\in \text{Band}}c^{*}_{\alpha}c^{b}_je^{-i(E_j^b-E_{\alpha})t\over\hbar} x^{*}_{\alpha,l}x_{j,l}+\text{c.c.},\nonumber
\end{eqnarray}
where $x_{j,l}^b$ and $x_{\alpha,l}$ are the excited-state probability amplitudes of the $l$th QE in the $j$th bound state and the $\alpha$ band state, respectively, see Eq. \eqref{eddgst}.
Both of the second and the third terms contain the oscillating frequencies $E_{\alpha}/\hbar$, which are continuously summed in the continuous energy band. Such terms tend to zero due to the out-of-phase interference of the different components in the long-time limit. Thus only the contributions of the bound states survive in the long-time limit. 

For the $N=2$ case, if one bound state is formed under the condition $|\Psi(0)\rangle=|e,g;\{0_\omega\}\rangle$, then we have $M=1$ and $c^b=x_1^{b\ast}$. Therefore, we obtain
\begin{eqnarray}
P_{1}(\infty ) &=&P_{2}(\infty )=|x_{1}^{b}|^{4}.
\end{eqnarray}
If two bound states are formed, then we have $M=2$ and $c^b_{j}=x^{b\ast}_{j,1}$. Thus, we obtain
\begin{eqnarray}
P_{1}(\infty )
&=&|x_{1,1}^{b}|^{4}+|x_{2,1}^{b}|^{4}+2|x_{1,1}^{b}x_{2,1}^{b}|^{2}\cos
({\Delta E^{b}t\over\hbar} ),~~~~~~~ \\
P_{2}(\infty )
&=&|x_{1,1}^{b}|^{4}+|x_{2,1}^{b}|^{4}-2|x_{1,1}^{b}x_{2,1}^{b}|^{2}\cos
({\Delta E^{b}t\over\hbar} ),
\end{eqnarray}
with $\Delta E^{b}=E_{1}^{b}-E_{2}^{b}$. These results match exactly with Eqs. \eqref{smsing} and \eqref{smtwo} directly derived from the Laplace transform.

The concurrence of two QEs is determined by $\mathcal{C}_{2}(t)=2|P_{1}(t)P_{2}(t)|^{1\over 2}$. We readily have
\begin{equation}\label{S30}
\mathcal{C}_{2}(\infty )=2|x_{1}^{b}|^{4}
\end{equation}
for $M=1$, and
\begin{eqnarray}\label{S31}
\mathcal{C}_{2}(\infty)&=&2\sqrt{[|x_{1,1}^{b}|^{4}+|x_{2,1}^{b}|^{4}]^2-4|x_{1,1}^{b}x_{2,1}^{b}|^{2}\cos^2
({\Delta E^{b}t\over\hbar} )}.\nonumber\\
\end{eqnarray}
for $M=2$. Equations \eqref{S30} and \eqref{S31} take the same form as the one of Eq. (6) in the main text.
\begin{figure}[tbp]
\centering
\includegraphics[width=\columnwidth]{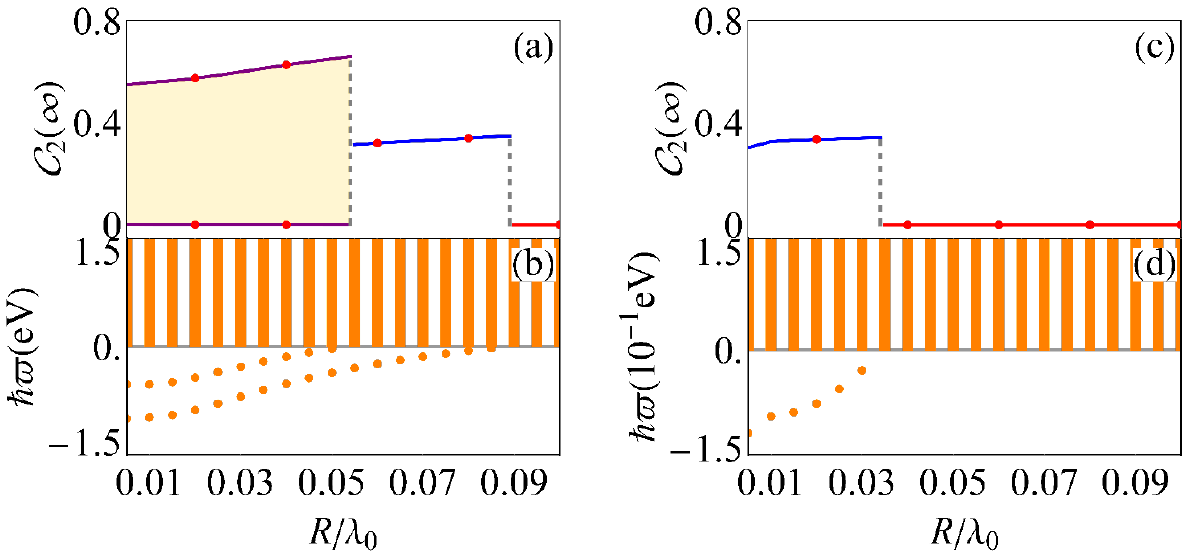}\\
\caption{$\mathcal{C}_{2}(\infty)$ and energy spectrum of the total system in different $R$ when $r_{a}=0.012\lambda_0$ in (a), (b) and $0.013\lambda_0$ in (c), (d). Other parameters are the same as Fig. 2(b) in the main text.}
\label{S1}
\end{figure}

The above proof clearly demonstrates the distinguish role of the bound states in determining the long-time steady state. The decoherence suppression caused by the formation of the bound states is guaranteed by the characters of the bound states as stationary states with isolated eigenenergies of the total system.

\section{Nanowire with different radius}\label{Appen-radiu}
The SPPs generally contain multiple modes labeled by $n=0,1,2,\cdot\cdot\cdot$, see Eq. \eqref{appsyn}, which have different properties in field confinement and propagation loss \cite{doi:10.1021/acs.chemrev.7b00441}. For example, the fundamental mode $n=0$ shows tighter confinement for small values of the nanowire radius $R$, but has a shorter propagation length due to higher absorption in the metal. In contrast, the confinement of the $n>0$ modes decreases and almost unconfined for the small $R$, but has larger propagation lengths. Thus, the nanowire with a proper $R$ is a key point for generating a strong QE-SPP coupling along a relative long propagation distance. In our scheme, the strong QE-SPP coupling is important in the generation of bound states. Thus we choose a small radius, i.e., $R=0.01\lambda_{0}\approx 15$ nm. In this case, it is the fundamental mode with $n=0$ that plays a major contribution to the QE-SPP interaction. If further reduce the radius, it is expected that the strengthened field confinement leads to the enhanced QE-SPP coupling, which is beneficial for the formation of bound states. However, the higher absorption would lead to a shortened distance $d$ between QEs in supporting the formation of the bound states. On the other hand, if we increase $R$, then the QE-SPP coupling is not strong enough to form the bound states, where, although the SPPs propagate a long distance, the entanglement still tends to zero in the steady state.

To see it more clearly, the energy spectrum and long-time limit of $\mathcal{C}_{2}(t)$ are plotted in Fig. \ref{S1} with the change of $R$ at two characteristic values of $r_a$. The related numerical results support the discussions above.

\bibliography{nanowire}

\end{document}